\documentclass[twocolumn,showpacs,aps,prl,superscriptaddress]{revtex4}

\usepackage{graphicx}
\usepackage{dcolumn}
\usepackage{amsmath}
\usepackage{epsfig}

\input pubboard/babarsym
\def\bpsikpm{\ensuremath{\Bpm \to \jpsi \Kpm}}
\def\bpsipipm{\ensuremath{\Bpm \to \jpsi \pipm}}

\def\BRpsipipm  {\ensuremath{\BR(\bpsipipm)}}
\def\BRpsikpm  {\ensuremath{\BR(\bpsikpm)}}

\def\psiee{\ensuremath{\jpsi \to e^+ e^-}}
\def\psimm{\ensuremath{\jpsi \to \mu^+ \mu^-}}
\def\psieehpm{\ensuremath{\jpsi(\epem) h^{\pm} }}
\def\psimmhpm{\ensuremath{\jpsi(\mumu) h^{\pm}}}

\def\hpm  {\ensuremath{h^{\pm}}\xspace}
\def\bpsieeh{\ensuremath{\Bpm \to \jpsi ( \to e^+ e^-) h^{\pm} }}
\def\bpsimmh{\ensuremath{\Bpm \to \jpsi ( \to \mu^+ \mu^-) h^{\pm} }}

\def\bupsih{\ensuremath{\Bu \to \jpsi h^{+} }}
\def\bmpsih{\ensuremath{\Bub \to \jpsi h^{-} }}
\def\bpsihpm{\ensuremath{\Bpm \to \jpsi h^{\pm} }}
\def\bpsipk{\ensuremath{\Bpm \to \jpsi \pipm (\Kpm) }}
\def\bxpsiX{\ensuremath{B \to \jpsi X}}

\def\psikpm{\ensuremath{\jpsi \Kpm}}
\def\psipipm{\ensuremath{\jpsi \pipm}}

\newcommand{\deltae}{\ensuremath{\Delta E}}
\newcommand{\deltaek}{\ensuremath{\Delta E_K}}

\newcommand{\deltaepi}{\ensuremath{\Delta E_\pi}}

\def\Dp      {\ensuremath{D^+}}
\def\Dm      {\ensuremath{D^-}}
\def\Dstar   {\ensuremath{D^{*}}}
\def\Dstarp  {\ensuremath{D^{*+}}}
\def\Dstarm  {\ensuremath{D^{*-}}}

\newcommand{\BABARPubYear}    {01}
\newcommand{\BABARPubNumber}  {14}
\newcommand{\SLACPubNumber} {8942}
\newcommand{\LANLNumber} {0108009}

\def\figurebox#1#2#3{%
    \def\arg{#3}%
    \ifx\arg\empty
    {\hfill\vbox{\hsize#2\hrule\hbox to #2{\vrule\hfill\vbox to #1{\hsize#2\vfill}\vrule}\hrule}\hfill}%
    \else
    {\hfill\epsfbox{#3}\hfill}%
    \fi}

\begin{document}

\begin{flushleft}
\babar-PUB-\BABARPubYear/\BABARPubNumber \\
SLAC-PUB-\SLACPubNumber\\
hep-ex/\LANLNumber\\[10mm]
\end{flushleft}

\title{
{\large \bf 
A Study of \bpsipipm\ and \bpsikpm\ Decays: Measurement
of the Ratio of Branching Fractions and Search for Direct 
\CP-Violating Charge Asymmetries} 
}

%
\author{B.~Aubert}
\author{D.~Boutigny}
\author{J.-M.~Gaillard}
\author{A.~Hicheur}
\author{Y.~Karyotakis}
\author{J.~P.~Lees}
\author{P.~Robbe}
\author{V.~Tisserand}
\affiliation{Laboratoire de Physique des Particules, F-74941 Annecy-le-Vieux, France }
\author{A.~Palano}
\author{A.~Pompili}
\affiliation{Universit\`a di Bari, Dipartimento di Fisica and INFN, I-70126 Bari, Italy }
% ...
\collaboration{The \babar\ Collaboration}
\noaffiliation

\date{\today}

\begin{abstract}
We have studied the \bpsipipm\ and \bpsikpm\ decays 
using a $20.7\invfb$ data set collected with the \babar\ detector. 
We observe a signal of $51\pm10$ \bpsipipm\ events and determine the 
ratio \BRpsipipm/\BRpsikpm\ to be 
$[3.91\pm 0.78 (stat.) \pm 0.19 (syst.)]\%$.
The \CP-violating charge asymmetries for the \bpsipipm\ and 
\bpsikpm\ decays are determined to be
$\calA_{\pi} = 0.01 \pm 0.22 (stat.) \pm 0.01 (syst.)$
and 
$\calA_K = 0.003 \pm 0.030 (stat.) \pm 0.004 (syst.)$.
\end{abstract}


\maketitle

The decay \bpsipipm\ is both Cabibbo-suppressed and
color-suppressed. If the leading-order tree diagram 
is the dominant contribution, its
branching fraction is expected to be about
5\% of the Cabibbo-allowed mode \bpsikpm. A comparable prediction can be
obtained with a simple model based on the factorization
hypothesis~\cite{NeuSte}.
Previous studies of this decay
were performed by the CLEO~\cite{CLEO} and CDF~\cite{CDF} collaborations.
Significant interference terms between the suppressed tree
and penguin amplitudes could produce a direct \CP-violating 
charge asymmetry in the \bpsipipm\ decays 
at the few percent level~\cite{Dunietz}.
On the contrary, a negligible direct \CP-violation is expected in the 
\bpsikpm\ decays, because for $\b\to\c\cbar\s$ transitions the Standard Model 
predicts that the leading- and higher-order diagrams are 
characterized by the same weak phase.

In this paper we present a measurement of the ratio of branching
fractions \BRpsipipm/\BRpsikpm\ along with a search for 
direct \CP-violation in these channels.
The data were recorded
at the \FourS\ resonance in 1999-2000 with the \babar\ detector at the \pep2
asymmetric-energy \epem collider at the Stanford Linear
Accelerator Center. 
The integrated luminosity is $20.7\invfb$, corresponding to $22.7$
million \BB\ pairs.
We fully reconstruct \bpsihpm\ decays, where $\hpm=\pipm, \Kpm$.
Signal yields and charge asymmetries 
are determined from an unbinned maximum likelihood fit
that exploits the kinematics of the decay to identify the 
\pipm, \Kpm\ and background components in the sample. This kinematic
separation is sufficiently good so that
no explicit particle
identification is required on the charged hadron \hpm, thereby
simplifying the analysis.
At the same time, particle identification can be used to perform 
a crosscheck of the measurement.

The \babar\ detector is described in detail elsewhere~\cite{nimpap}.
A five-layer silicon vertex tracker (SVT) and a 40-layer drift
chamber (DCH), in a 1.5-T solenoidal magnetic field, provide detection
of charged particles and measurement of their momenta. The transverse
momentum resolution is $\sigma_{p_{t}}/p_{t} = (0.13 \pm
0.01)\% \cdot  p_{t} + (0.45 \pm 0.03) \% $, where $p_{t}$ is measured
in \gevc. 
Electrons are detected in a CsI electromagnetic calorimeter (EMC),
while muons are identified in the magnetic flux return system (IFR), 
which is instrumented with
multiple layers of resistive plate chambers. A ring-imaging Cherenkov
detector (DIRC) with a quartz bar radiator provides charged
particle identification. 

An electron candidate is selected according to the ratio of the 
energy detected in the EMC to track momentum, the cluster shape in the
EMC, the energy loss in the DCH, and the DIRC Cherenkov angle, if
available.
A muon candidate is selected according to the difference between the
expected and measured thickness of absorber traversed,
the match of the hits in the IFR with the extrapolated track, the
average and spread in the number of hits per IFR layer, and the
energy detected in the EMC.

\psimm\ candidates are constructed from two identified muons with polar
angle in the range $[0.3,2.7]$ radians and with invariant mass $
3.06 < M_{\mumu} < 3.14 $\gevcc. The absolute value of the cosine of the
helicity angle of
the \jpsi decay is required to be less than 0.9. \psiee\ candidates are
constructed from two identified electrons with polar
angle in the range $[0.41,2.409]$ radians and with invariant mass $ 2.95
< M_{\epem} < 3.14 $\gevcc. The absolute value of the cosine of the
helicity angle is
required to be less than 0.8.  

\begin{figure}[!htb]
\begin{center}
  \includegraphics[width=\linewidth]{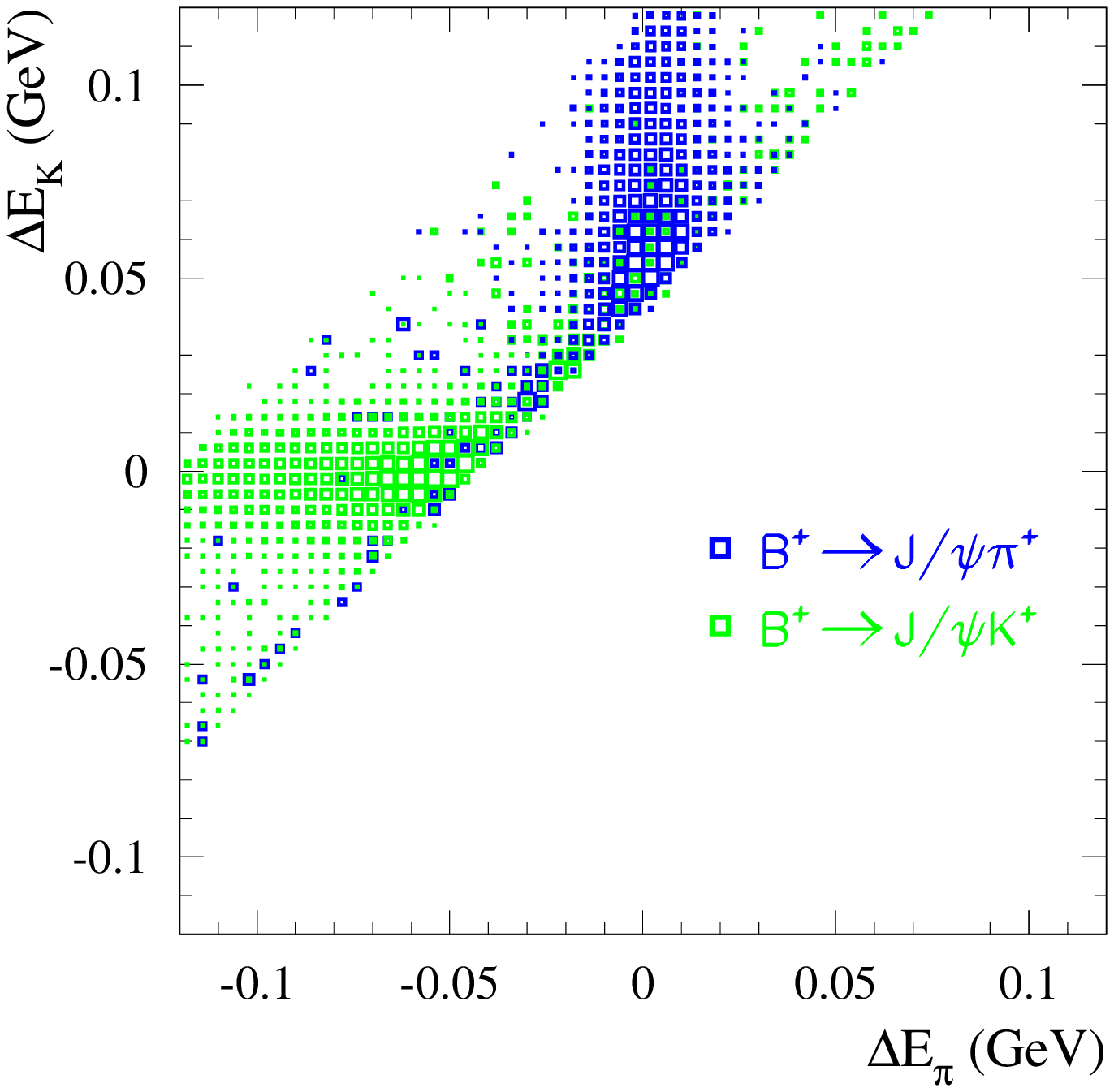}
  \caption{Distribution of \deltaek\ vs. \deltaepi\ for \bpsikpm\
    and \bpsipipm\ events from Monte Carlo simulations. 
    }
\label{fig:devar}
\end{center}
\end{figure}

\Bpm\ candidates are formed from the combination of a reconstructed
\jpsi, constrained to the world average mass~\cite{PDG2000}, and a
charged track \hpm. A vertex constraint is applied to the
reconstructed tracks before computing two
kinematic quantities of the \Bpm\ candidate used to discriminate
signal from background. 
We define the beam energy-substituted mass \mes\ as
\begin{equation}
\mes =  \sqrt{[ (s/2+{\bf p_i}
      \cdot{\bf p_\B})^2 / E_i^2 ]
  - |{\bf p_\B}|^2} \, ,
\end{equation}
where $\sqrt{s}$ is the total energy of the \epem\ system in the
\FourS\ rest frame, and $(E_i, {\bf p_i})$ and $(E_\B, {\bf p_\B})$ 
are the four-momenta of the \epem\ system and the reconstructed 
\B\ candidate, both in the laboratory frame. 
We define the kinematic variable \deltaepi\ (\deltaek) as 
the difference between the reconstructed energy of 
the \Bpm\ candidate and the beam
energy in the \FourS\ rest frame assuming $\hpm=\pipm(\Kpm)$.
We require $\vert \deltaepi \vert < 120 \mev $, $\vert
\deltaek \vert < 120 \mev  $ and $ \mes > 5.2 \gevcc$. 
Figure~\ref{fig:devar} shows the distribution for 
Monte Carlo simulations of
\bpsipipm\ and \bpsikpm\ events in the $(\deltaepi,\deltaek)$ plane.

\begin{figure}[!htb]
\begin{center}
  \includegraphics[width=\linewidth]{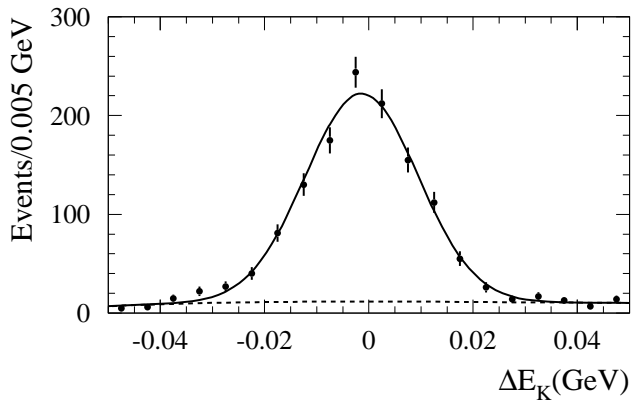}
  \caption{The \deltaek\ distribution and fit for the
    events in the data
    sample with $\mes > 5.27 \gevcc$. The dashed curve represents 
    the background contribution.
    }
\label{fig:desam2in1}
\end{center}
\end{figure}

The selected sample contains $1074$
\bpsimmh\ and $1081$ \bpsieeh\ candidates. 
A fit to the \deltaek\ distribution with the sum of a Gaussian and
a polynomial function, modeling the \bpsikpm\ signal and the background
contribution, is shown in Fig.~\ref{fig:desam2in1}.

\begin{figure}[!htb]
\begin{center}
  \includegraphics[width=\linewidth]{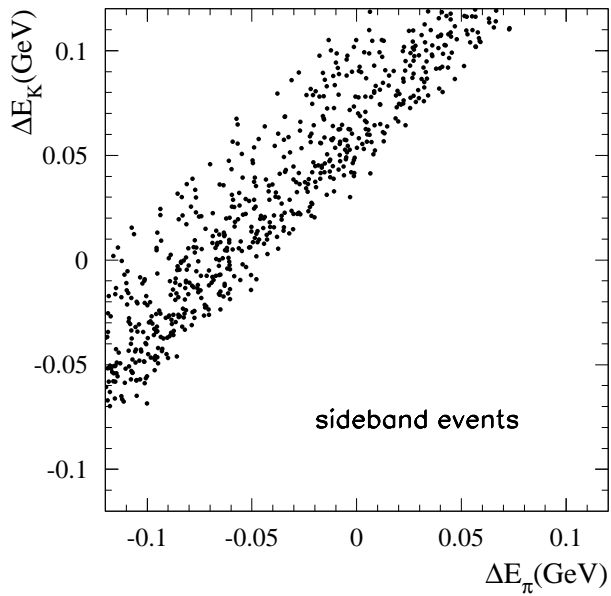}
  \caption{Distribution of \deltaek\ vs. \deltaepi\ for the events
    in the \mes\ sideband of the data sample.
    }
\label{fig:bkd}
\end{center}
\end{figure}

\begin{figure}[!htb]
\begin{center}
  \includegraphics[width=\linewidth]{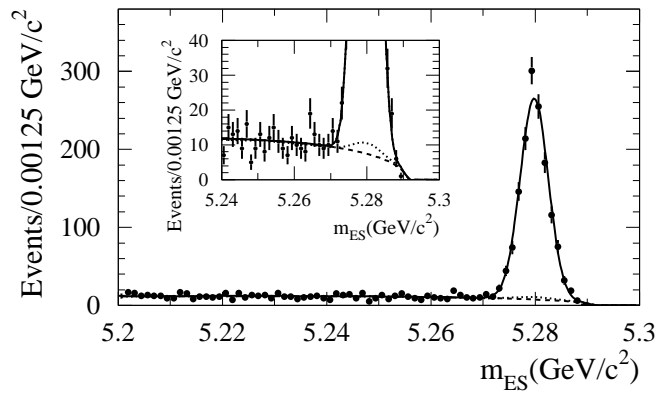}
  \caption{The \mes\ distribution and fit for the events in the data
    sample. The ARGUS (dashed curve) and peaking (dotted curve)
    components of the background are also displayed.
    }
\label{fig:sample}
\end{center}
\end{figure}

The background contaminating the sample is characterized 
with events in the data that are sufficiently far from the
typical signal regions (sidebands of
the data sample). We define \mes\ sideband events by the requirement that
$ 5.2 < \mes < M_\B -4\sigma(\mes)=5.27\gevcc$, where $M_\B$ is the 
world average \Bpm\ mass~\cite{PDG2000} and $\sigma(\mes)$ is the 
\mes resolution; their distribution in the
$(\deltaepi,\deltaek)$ plane is shown in
Fig.~\ref{fig:bkd}. We define
\deltaek\ and \deltaepi\ sideband events by the requirement that
$ 120 > \vert \deltaek \vert > 4 \sigma(\Delta E) = 42 \mev$ and
$ 120 > \vert \deltaepi \vert > 4 \sigma(\Delta E) = 42 \mev$,
where $\sigma(\Delta E)$ is the
width of the fitted Gaussian in Fig.~\ref{fig:desam2in1}. 
The distribution in \mes\ of the sideband events is modeled by an ARGUS
function~\cite{ARGUS_bkgd}, with an additional Gaussian peak in the
\mes\ signal region for
events from other \bxpsiX\ decays. The number of
background events in this peak 
has been estimated to be $10\pm4$ 
with detailed Monte Carlo simulation of
inclusive charmonium decays.
Figure~\ref{fig:sample} shows the \mes\ distribution for the data
sample, along with the fit.

Our fit to the data sample is based on maximizing the following 
extended likelihood
function:
\begin{equation}
L = {\rm e}^{-\sum_{i}N_i}
\prod_{j=1}^M \sum_{i} P_{i}(\deltaepi^j, p^j, \mes^j) N_i \, ,
\label{eq:likedef}
\end{equation}
where $j$ is the index of the event, $i$
is the index of the hypothesis ($i=\pi,K,bkd$), $N_i$ are the yields
for the \bpsipipm, \bpsikpm, and background events in the sample, and
$M$ is the total number of events.
The observables 
\deltaepi, the momentum $p$ of the final-state charged
hadron computed in the laboratory frame, and \mes\ are 
used as arguments of the probability density functions (PDF) $P_{i}$.
The PDFs are mainly determined from data with limited input from simulation.

It is useful to define the new variables $ D = \deltaek -
\deltaepi = \gamma \left( \sqrt{p^2 + m_K^2}-\sqrt{p^2 + m_\pi^2}
\right)$, where $\gamma$ is the Lorentz boost from the laboratory 
frame to the \FourS\
rest frame, and $S = \deltaek + \deltaepi =
2 \deltaepi + D$. These variables have the property that $(\deltaepi,D)$ in
the pion hypothesis, $(\deltaek,D)$ in the kaon hypothesis, and
$(S,D)$ in the background hypothesis are uncorrelated
at the 1\% level. Therefore, with appropriate
transformations of variables, each $P_{i}(\deltaepi, p, \mes)$ can be
written as a product of one-dimensional PDFs:
\begin{eqnarray}
P_{\pi}(\deltaepi, p, \mes ) &=& f_\pi(\deltaepi) g_\pi(D) h_\pi(\mes)
\, , \\
P_{K}(\deltaepi, p, \mes ) &=& f_K(\deltaek) g_K(D) h_K(\mes) \, , \\
P_{bkd}(\deltaepi, p, \mes ) &=& f_{bkd}(S) g_{bkd}(D) h_{bkd}(\mes).
\end{eqnarray}

The $f_\pi(\deltaepi)$, $f_K(\deltaek)$, $h_\pi(\mes)$, and
$h_K(\mes)$ components
are the $\Delta E$ and \mes\ resolution functions for the signals. The
mean values and the Gaussian widths are allowed to float as free
parameters in the likelihood
fit and are extracted together with the yields. This strategy reduces
the systematic error due to possible inaccuracies of the \deltae\ and
\mes\ description in Monte Carlo simulations.
 
\begin{figure}[!htb]
\begin{center}
  \includegraphics[width=\linewidth]{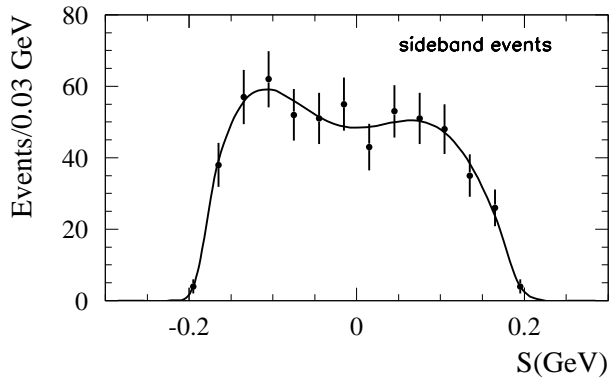}
  \caption{The $S$ distribution and fit for the events in the \mes\
    sideband of the data sample. 
    }
\label{fig:Spdf}
\end{center}
\end{figure}

The $f_{bkd}$ component is represented by a phenomenological
function with eight fixed parameters, all estimated from the
distribution of $S$ for the events in the \mes\ sideband 
(Fig.~\ref{fig:Spdf}). 

The $h_{bkd}$ component is represented
by the sum of an ARGUS and a Gaussian
function, with parameters estimated from the distribution of \mes\ for the
events in the \deltaek\ and \deltaepi\ sidebands.

The $g$ components are each
represented by a phenomenological function with seven fixed
parameters. The parameters
are estimated with Monte Carlo simulations 
for the $\pi$ and $K$ hypotheses, and
with events in the \mes\ sideband for the background
case.
A comparison of the $D$ distributions in the three hypotheses shows
that this variable, introduced by our procedure for 
factorizing PDFs, 
provides little discriminating power. 

From the maximum likelihood fit to the selected sample
we obtain $N_{\pi} = 52 \pm 10$, $N_{K} = 1284 \pm 37$, and $N_{bkd} =
819 \pm 31$.
The correlation coefficient between $N_{\pi}$ and $N_{K}$ is $-0.04$.
The confidence
level of the fit, defined as the probability to obtain a
maximum value of the likelihood smaller than the observed value, is
$54 \%$, estimated by Monte Carlo techniques.
The statistical significance of the \bpsipipm\ signal, evaluated 
from the change in the maximum value of $\ln L$ when we
constrain $N_{\pi}=0$, is $7.0 \sigma$.

\begin{figure}[!htb]
\begin{center}
  \includegraphics[width=\linewidth]{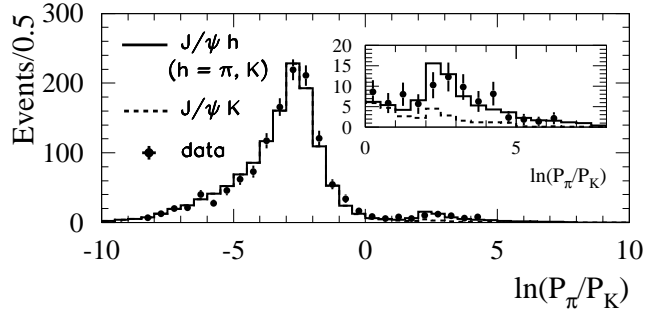}
  \caption{The $\ln(P_{\pi} / P_{K})$ distribution for
events in the data sample (after the subtraction of the background
  component in each bin) and from Monte Carlo simulations of \bpsipk\
  events; the distributions are
  normalized to the yields extracted from the maximum likelihood fit.
    }
\label{fig:loglike}
\end{center}
\end{figure}

\begin{figure}[!htb]
\begin{center}
  \includegraphics[width=\linewidth]{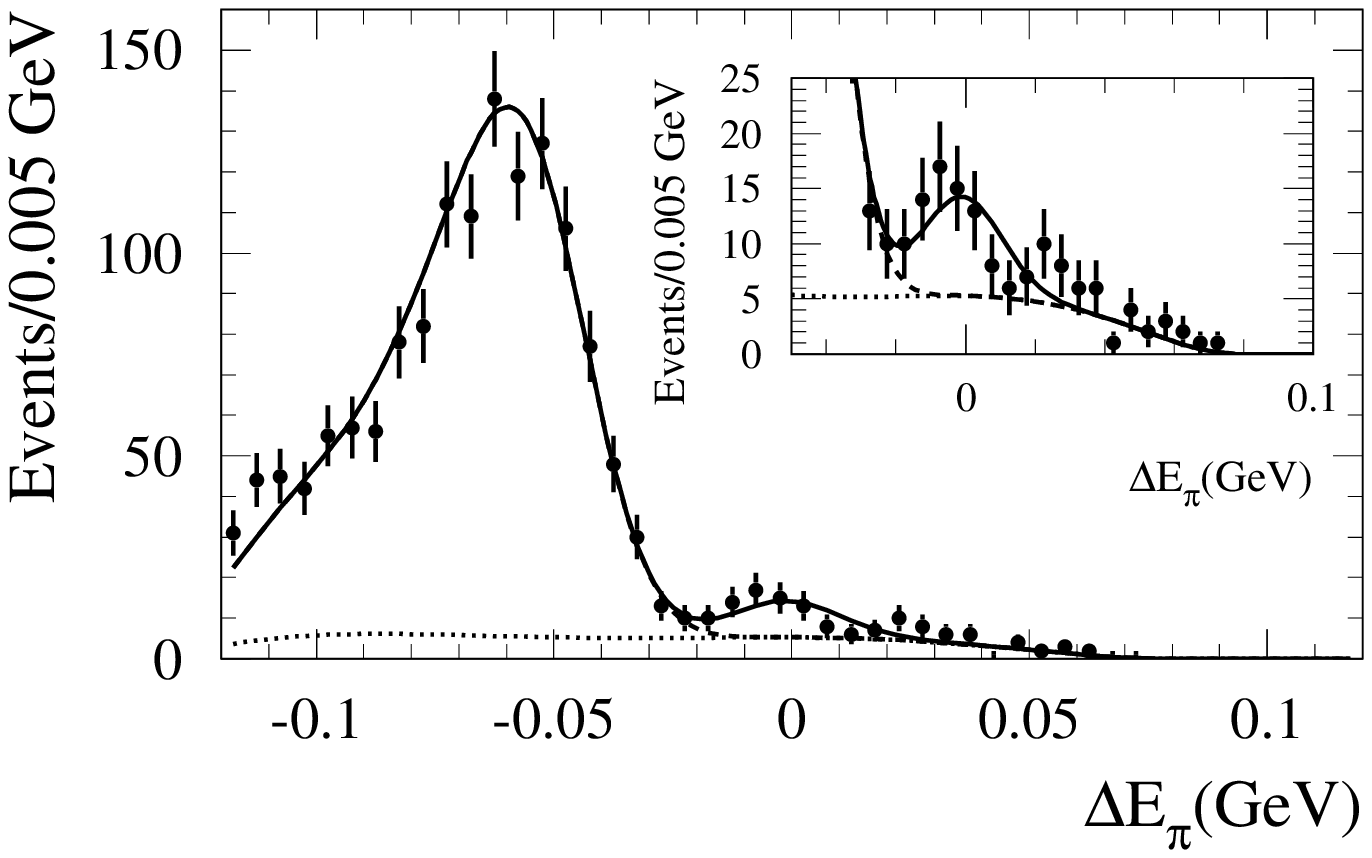}
  \caption{The \deltaepi\ distribution for events with $\mes > 5.27
    \gevcc$ compared with the fit result (solid curve).
    The dotted curve represents the fitted contribution 
    from the background alone, while the dashed curve represents 
    the fitted contributions from the sum of background and \psikpm\ 
    components. 
    The PDFs of the \deltaepi\ variable in the \psikpm\ and 
    background hypotheses
    have been obtained with a numerical integration of the 
    $P_i$ PDFs: $p_K(\deltaepi)=\int{f_K(x)
        g_K(x - \deltaepi) \, {\mathrm d}x }$, 
      $p_{bkd}(\deltaepi)=\int{f_{bkd}(x + \deltaepi) g_{bkd}(x - \deltaepi)\, 
    {\mathrm d}x }$. 
    }
\label{fig:depifit}
\end{center}
\end{figure}

The distribution of $\ln(P_{\pi} / P_{K})$ for the sample, after
subtraction of the background component in each bin, is shown in
Fig.~\ref{fig:loglike}.
The background distribution is normalized to the number of 
background events from the fit.
The distribution of $\ln(P_{\pi} / P_{K})$ for simulated
signal samples, normalized to the yields extracted
from the likelihood fit, is also shown.
The distribution in
\deltaepi\ for the events in the data sample with 
$\mes > 5.27 \gevcc$ is shown in
Fig.~\ref{fig:depifit}, along with the likelihood 
fit result.
 
Possible biases in the fitting procedure were investigated by
performing the fit on simulated samples of known
composition and of the same size as the data. The differences,
$\Delta_{\pi}$ 
and $\Delta_{K}$, between
the extracted and the input values are consistent with 0. However we
correct the yields for the observed deviations 
$\Delta_{\pi}=1.1 \pm 2.2$ and $\Delta_{K} = -11.3 \pm 8.8$.
The corrected yields are $51 \pm 10$ and $1296 \pm 38$ for 
\psipipm\ and \psikpm, respectively.

The use of particle identification for the charged hadron \hpm\ has been
investigated by adding to
the likelihood, as an additional argument, the Cherenkov angle
$\theta_C$ measured in the DIRC for this track.
The PDFs for the variable $\theta_C$ are
determined from data and parameterized as Gaussian functions, with
mean values and widths that depend on the momentum of the track.
A fit with a modified likelihood function is performed with the
subsample of events where the particle identification information is
available. 
The ratio of branching fractions is determined separately for the
\psimmhpm\ and \psieehpm\ samples. A detailed comparison, reported in 
Table~\ref{tab:compar12}, shows
that the addition of particle identification does not significantly 
change the statistical precision of the results, which 
are consistent to within $1.6\sigma$.

Based on the fitted event yields, we find the
ratio of branching fractions to be
\begin{equation*}
\frac{\BRpsipipm}{\BRpsikpm} = [3.91\pm 0.78 (stat.) \pm 0.19 (syst.)]\%.
\end{equation*}
The dominant systematic error ($0.17 \%$) comes from the 
uncertainty in the correction factors, $\Delta_{\pi}$
and $\Delta_{K}$, due to the limited statistics of the simulated samples.
The uncertainty in the fixed parameters of the PDFs, determined 
by fits to simulated or non-signal data sets, affects several 
aspects of the likelihood fit: the characterization of the $S$ and $D$ 
distributions; the characterization of the
\mes distribution for the background (including the fraction 
of peaking background events); and the fraction of
signal events in the tails of the $\Delta E$ distribution.
This uncertainty contributes $0.07\%$ to the systematic error.
Contributions due to any possible difference in the
reconstruction efficiencies for \psipipm\ and
\psikpm\ events are found to be negligible, 
as are uncertainties due to inaccuracies in the description of the tails
of the $\Delta E$ resolution function. 

\begin{table}[!htb]
\begin{center}
  \caption{Measurements of \BRpsipipm/\BRpsikpm\ obtained
    with the original (fit 1) and a modified likelihood
    function (fit 2) that includes particle identification for \hpm. 
    The error on the difference $\Delta$ between the two 
    measurements is estimated as $\sigma_{\Delta} = 
    \sqrt{ | \sigma_1^2 - \sigma_2^2} | $.
}
\label{tab:compar12}
\begin{tabular}{lccc}
\hline\hline                
sample & fit 1 & fit 2 & $\Delta/\sigma_{\Delta}$ \\ \hline
\psimmhpm & $(4.2 \pm 1.0)\% $ & $(4.7 \pm 1.1)\% $ & 1.1 \\ \hline
\psieehpm & $(3.5 \pm 1.2)\% $ & $(4.1 \pm 1.3)\% $ & 1.2 \\
\hline\hline                
\end{tabular}
\end{center}
\end{table}

Our determination of the ratio of branching fractions is consistent with
the expectation reported in~\cite{NeuSte} and with 
previous measurements~\cite{CLEO, CDF},
but has a substantially lower uncertainty than 
the world average value of $(5.1 \pm 1.4)\%$~\cite{PDG2000}.

To study direct \CP-violation in these channels, 
we modify the likelihood function in 
Eq.~\ref{eq:likedef} as follows:
\begin{equation}
L^{\prime} = {\rm e}^{-\sum_{i}N_i}
\prod_{j=1}^M \sum_{i} P^{\prime}_{i}
(\deltaepi^j, p^j, \mes^j, q^j) N_i\, ,
\label{eq:likedefq}
\end{equation} 
where $q$ is the charge of $h^\pm$.
We factorize the PDFs as
\begin{equation}
P^{\prime}_{i} (\deltaepi, p, \mes, q) = P_i(\deltaepi, p, \mes) c_i(q)\, ,
\end{equation} 
where $c_i(q)$ is the probability for the final state 
charged hadron, in a certain hypothesis, to have charge $q$.
The $c_i$ can
be written in terms of the \CP-violating charge asymmetries 
$\calA_i$ as
\begin{equation}
c_i(q) =  \frac{1}{2}[(1-\calA_{i})f^{+}(q) + (1+\calA_{i})f^{-}(q)] \, ,
\end{equation}
where
\begin{eqnarray}
\calA_{i} & = & \frac{N_i^- - N_i^+}{N_i^- + N_i^+} \, , \\
f^{+}(q) & = & 
  \left\{ 
        \begin{array}{ll}
        1 & \mbox{if $q= +1$}\\
        0 & \mbox{if $q= -1$} \, ,
        \end{array}     
           \right. \\
f^{-}(q) & = & 
  \left\{ 
        \begin{array}{ll}
        0 & \mbox{if $q= +1$}\\
        1 & \mbox{if $q= -1$} \, .
        \end{array}     
           \right. 
\end{eqnarray} 
The asymmetry observables $\calA_i$ are allowed to float 
as free parameters in the likelihood fit and are extracted together
with the yields.

We impose additional requirements on 
the charged track \hpm\ in the events to 
be used in the fit, selecting only those 
tracks for which the tracking efficiency has been 
accurately measured from data.
Tracks are required to have a polar angle in the 
range $[0.41,2.54]$ radians, to include 
at least $12$ DCH hits, to have $p_t > 100 \mevc$, and 
to point back to the nominal interaction point 
within $1.5 \cm$ in the vertical plane and within 
$3 \cm$ along the longitudinal direction. 
The selected sample contains $982$ \bmpsih\
and $970$ \bupsih\ candidates. 

From the maximum likelihood fit to the data sample 
we obtain $\calA_{\pi}=0.01 \pm 0.22$, 
$\calA_{K}=-0.001 \pm 0.030$, $\calA_{bkd}=0.018 \pm 0.039$.
The correlation coefficient between $\calA_{\pi}$ and 
$\calA_{K}$ is $-0.03$.

The uncertainty in the fixed parameters of the PDFs, determined 
by fits to simulated or non-signal data sets, contributes 
$0.0056$ and $0.0002$  to the systematic 
error on $\calA_{\pi}$ and $\calA_K$, respectively.
The difference in tracking efficiency between positively 
and negatively charged tracks -- primarily pions -- 
has been studied in hadronic events by comparing the independent 
SVT and DCH tracking systems.
The corrections to the asymmetries $\calA_{\pi}$ and $\calA_K$ 
are negligible. The uncertainty on the corrections contributes 0.0026
and 0.0020 to the systematic error on $\calA_{\pi}$ and $\calA_K$, 
respectively. 
The fake asymmetry due to the different probability
of interaction of \Kp\ and \Km\ in the detector material
before the DCH is estimated to be $- 0.0039$. 
We correct $\calA_K$ for this quantity and conservatively assume 
a contribution of $0.0039$ to the systematic uncertainty.
This represents the dominant systematic error on $\calA_K$.
A more careful evaluation of the materials and of \Kp/\Km\
cross-section differences will make it possible to substantially 
reduce this contribution.

We determine the \CP-violating charge asymmetries to be
\begin{eqnarray*}
\calA_\pi &=&  0.01  \pm 0.22 (stat.) \pm 0.01 (syst.) \, , \\
\calA_K   &=&  0.003 \pm 0.030 (stat.) \pm 0.004 (syst.) \, .
\end{eqnarray*}
These results are consistent with Standard Model expectations 
and with the measurement reported in~\cite{CLEOdirect}.

As a crosscheck, $\calA_K$ has been determined also with a 
simple analysis based on the counting of \bpsikpm\ signal 
events in the \mes\ peak. 
The result is compatible with 
the likelihood fit analysis:
$A_K = 0.005\pm 0.030 (stat.) \pm 0.004 (syst.)$.

We observe no evidence for \CP-violation in \bpsipipm\ 
or \bpsikpm\ decays. 
These results are statistically limited and can 
be expected to improve with additional data.

We are grateful for the excellent luminosity and machine conditions
provided by our \pep2\ colleagues.
The collaborating institutions wish to thank 
SLAC for its support and kind hospitality. 
This work is supported by
DOE
and NSF (USA),
NSERC (Canada),
IHEP (China),
CEA and
CNRS-IN2P3
(France),
BMBF
(Germany),
INFN (Italy),
NFR (Norway),
MIST (Russia), and
PPARC (United Kingdom). 
Individuals have received support from the Swiss NSF, 
A.~P.~Sloan Foundation, 
Research Corporation,
and Alexander von Humboldt Foundation.

\end{document}